\documentclass[12pt,preprint]{aastex}

\begin{document}

\title{The globular cluster NGC 1978 in the Large Magellanic Cloud}

\author{Alessio Mucciarelli}
\affil{Dipartimento di Astronomia, Universit\`a 
degli Studi di Bologna, Via Ranzani, 1 - 40127
Bologna, ITALY}
\email{alessio.mucciarelli@studio.unibo.it}

\author{ Francesco R. Ferraro}
\affil{Dipartimento di Astronomia, Universit\`a 
degli Studi di Bologna, Via Ranzani, 1 - 40127
Bologna, ITALY}
\email{francesco.ferraro3@unibo.it}

\author{Livia Origlia}
\affil{INAF - Osservatorio Astronomico di Bologna, Via Ranzani, 1 - 40127
Bologna, ITALY}
\email{livia.origlia@bo.astro.it}

\author{Flavio Fusi Pecci}
\affil{INAF - Osservatorio Astronomico di Bologna, Via Ranzani, 1 - 40127
Bologna, ITALY}
\email{flavio.fusipecci@oabo.inaf.it}

\begin{abstract}

We have used deep high-resolution Hubble Space Telescope ACS observations 
to image the cluster NGC 1978 in the Large Magellanic Cloud. 
This high-quality photometric data set allowed us to confirm 
the high ellipticity ($\epsilon\sim$0.30$\pm$0.02) of this stellar system.
The derived Color Magnitude Diagram allowed a detailed study of the main evolutionary 
sequences: in particular we have detected the so called Bump along the Red Giant Branch
(at $V_{555}$=19.10$\pm$0.10). This is the first detection of this feature in an 
intermediate-age cluster.\\
Moreover the morphology of the evolutionary sequence and their population ratios 
have been compared with the expectations of different theoretical models 
(namely BaSTI, PEL and Padua) in order to quantify the effect of convective 
overshooting. The best agreement (both in terms of morphology and star counts) 
has been found the PEL ({\sl Pisa Evolutionary Library}) 
isochrone with Z=0.008 
(consistenly with the most recent determination of the cluster metallicity, 
[M/H]=-0.37 dex)
and a mild overshooting efficiency
($\Lambda_{os}$=0.1). By adopting this theoretical set 
 an age of $\tau$=1.9$\pm$0.1 Gyr has been obtained.

\end{abstract}  
 
\keywords{Magellanic Clouds --- globular clusters: individual (NGC~1978) ---
techniques: photometry}   

\section{Introduction}   
\label{intro}

The Large Magellanic Cloud (LMC) is the nearest galaxy of the Local Group with 
evident star formation activity. The LMC has a very popolous system of globular
clusters that covers a large range of metallicities and ages (see \citet{ols96} 
and references therein). We can distinguish an
old population, coeval with the Galactic globular cluster (GGC) system, an
intermediate (1-3 Gyr) and a young population, ranging
from a few million years to 1 Gyr. 
Hence the stellar populations of the LMC cluster system represent an important 
{\sl laboratory} to study and test stellar evolution models in a 
different age-regime with respect to the Galactic system.

The first surveys devoted to study the properties of these clusters have 
been carried out
in the optical \citep{vdb81} and in the near-infrared \citep{persson83}
spectral range and were
based on measurements of their integrated colours \citep{swb, ef85}. 
However, only in the last decade the main properties of 
the resolved stellar populations in LMC globulars have been investigated 
by means of high resolution photometry of individual stars
\citep{valle94, c94, f95, broc01, mac03, f04, m06}.
These works
have produced a number of interesting results:
({\sl i})  the oldest clusters of the LMC 
are coeval with GGCs \citep{testa95, brocato96, olsen98, mac04}. 
The recent {\sl census} of the old LMC globulars presented by \citet{mac04} 
counts a total of 15 objects with ages $\ge$ 10 Gyr;
({\sl ii}) the dominant population is constituted of $\sim$100 young and 
intermediate age clusters,
with ages $<$3 Gyr;
({\sl iii}) the absence of objects in the huge age range from 
$\sim$3 to $\sim$ 13 Gyr (the so-called {\sl Age Gap}):
The {\sl Age Gap} problem has been largely 
investigated \citep[see i.e.][]{bekki} 
and clusters in the {\sl Age Gap} actively searched \citep{geisler97,rich01,mac06}, 
however 
up to date, only
one candidate, namely ESO 121-SC03, possibly falls in such an age range.
In spite of these efforts, the overall formation history of the LMC is 
still an open issue,
since a firm age-metallicity relation (AMR) has not been established, yet.
Indeed, homogeneous and accurate metallicities and ages 
for a significant sample of LMC clusters
are mandatory to derive a reliable AMR.

Accurate ages from the measurement of the Main-Sequence (MS) Turn-Off (TO) 
region are still sparse and very model (i.e. isochrones) dependent.
The only homogeneous age-scale available still relies 
on the so called s-parameter \citep{ef85, ef88}, 
an empirical quantity related to the position of the cluster in the 
dereddened (U-B) vs (B-V) color-color diagram. This 
parameter linearly correlates with the logarithm of the age \citep{ef85, gir95}. 

Most chemical abundance determinations for LMC clusters are based on 
low-resolution spectra \citep{ols91}, integrated infrared 
spectroscopy \citep{oliva98} 
or photometric techniques \citep{dirsch, larsen00}. Detailed chemical 
abundances from 
high-resolution spectra are still sparse \citep{hill00, j06, f06}.

With the ultimate goal of constructing a homogeneous age-metallicity scale 
for the LMC clusters,
we started a program which makes use of the last generation of 
instruments (imager and multi-object spectrograph) in order to 
perform an appropriate study of stellar population, age, metal content 
and structural parameters for a number of pillar clusters.
In this paper we present the results for 
one of the most massive and luminous globulars in the LMC, namely NGC 1978.
This cluster has been considered for years a peculiar object ({\sl i}) 
because of its high-ellipticity \citep[$\epsilon$=0.3,][]{fischer},
significantly larger than the one typically measured in stellar clusters  
 \citep{goodwin}, and ({\sl ii}) because it has been suspected to 
 harbour a chemically 
 inhomogeneous stellar population \citep{alcaino, hill00}. 
 However, \citet{f06} has recently presented high resolution 
spectra for eleven giants in this cluster, obtaining [Fe/H]=-0.38 dex,
 with a very low dispersion ($\sigma$=0.07 dex), firmly excluding 
the presence of a significant metallicity spread in this cluster.
Here we present a high precision CMD of the cluster based on observations 
obtained with the Advanced Camera for Surveys (ACS) on board 
the Hubble Space Telescope (HST). The morphology and the 
population ratios along the main evolutionary sequences have been 
used to quantify the effect of convective overshooting. 
Finally, by combining the precise measure of the MS-TO with 
accurate estimate of the cluster metallicity \citep{f06}
we also provide an accurate estimate of the cluster age.

\section{Observations and data analysis}

\subsection{Observations}
The photometric dataset consists of high-resolution images obtained 
with ACS@HST (300 sec and 200 sec long) 
through the F555W and F814W filters. These images have been
retrieved from ESO/ST-ECF Science Archive (Proposal ID 9891, Cycle 12).
The observations have been obtained with the Wide Field Channel (WFC) that 
provides a field of view of $\approx$200'' $\times$200'' with a
 plate scale of 0.05 
arcsec/pixel. The WFC is a mosaic of two CCDs, both with 
4096$\times$2048 pixels
separated by a gap of $\sim$50 pixels.
The first chip has been centered on the cluster center, while the 
second chip sampled a contiguos field. 
All images were reduced with the ACS/WFC pipeline, in order  to perform  
bias and dark subtractions and flatfield correction.
The photometric reduction was performed by using the {\it DAOPHOT-II} \citep{stet} 
Point Spread Function (PSF) fitting method.
The output catalog includes about 40,000 objects from the first chip and more
 than 10,000 from the second 
chip and it has been calibrated in the {\sl ACS/WFC Vega-mag} system, following
the prescriptions of \citet{bedin05}. Finally, the ACS catalog has been 
astrometrized in the 2MASS 
astrometric system by matching the IR catalog presented in \citet{m06}.

\subsection{The Color-Magnitude Diagram}

Fig. 1 shows the calibrated color-magnitude diagram (CMD) for the chip centered 
on the cluster. 
Stars in the brightest portion of the Giant Branches could be saturated and/or in
 the regime of non linearity of the CMD. Hence for stars brighter than $V_{555}$=17.6 
 (this magnitude level is marked with an horizontal dashed line in Fig. 1), magnitudes,  
 colours and level of incompleteness are not safely measured.  
This CMD (reaching the magnitude limit of  $V_{555}\sim$26)
shows the typical evolutionary features of an intermediate-age stellar population,
 namely:

{\it (1)}  the brightest portion of the MS at $V_{555}<$21 
shows a hook-like shape, typical of the evolution of 
intermediate-mass stars ( $M>1.2M_{\odot}$ ) which develop a convective core
\footnote{ Note that the width of the distribution in color of the bright portion 
of the MS ($\sigma_{(V-I)}\sim$0.05 mag) turns out to be fully consistent with the 
observational errors estimated from the completeness experiments 
($\sigma_{V}\sim\sigma_{I}\sim$0.03 mag, corresponding to $\sigma_{(V-I)}\sim$0.04 mag).}.
In particular, the so-called {\sl overall contraction} phase \citep{sc}
is clearly visible between the brightest portion of the MS and the 
beginning of the 
Sub-Giant Branch (SGB) at $V_{555}\sim$20.9.

{\it (2)} the SGB is a narrow, well-defined sequence at  
$V_{555}\sim$20.7, with a large extension in color ($\delta(V_{555}$-$I_{814})$
$\sim$0.6 mag). The blue edge of the SGB is broad
and probably affected by blending, especially in the most internal region of 
the cluster.
 
{\it (3)} the Red Giant Branch (RGB) is fully populated;
 this is not surprising since this cluster has 
already experienced the RGB Phase Transition 
(see the discussion in \citet{f04} and \citet{m06}).

{\it (4)}  the He-Clump is located at $V_{555}\sim$19.1 and 
$(V_{555}-I_{814})\sim$1.15.

Fig. 2 shows the CMD of the 
external part of the ACS@HST field of view (corresponding to 
r$>$140'' from the cluster center). This CMD
can be assumed as representative of the 
field population surrounding the cluster. In particular, the CMD shows 
two main components:

{\it (1)} a blue sequence extended up to $V_{555}\sim$17.

{\it (2)} a SGB which merges into the MS 
at $V_{555}\sim$22.2, corresponding to a population of $\approx$5 Gyr.
 We interpret this feature as 
a signature of the major star-formation episode occurred 
5-6 Gyr ago, when LMC and Small Magellanic Cloud (SMC) were 
gravitationally bounded \citep{bekki}. 
 
\subsection{Completeness}

In order to quantify the degree of completeness of the final 
photometric catalog, we 
used the well-know artificial star technique \citep{mateo}, and  
simulated a population of stars in the same magnitude range 
covered by the observed CMD (excluding stars brighter than 
$V_{555}$=17.6, corresponding to the saturation level)
and with $(V_{555}-I_{814})\sim$0.8 
mean color.
The artificial stars have been added to the original images and
the entire data reduction procedure has been repeated using
the {\sl enriched} images.
The number of artificial stars simulated in each run ($\sim$ 2,000) 
are a small percentage ($\sim$5\%)
of the detected stars, hence they cannot
alter the original crowding conditions. A total of $\sim$250 runs 
were performed and more than 500,000 stars have been simulated.
In order to minimize 
the effect of incompleteness correction, we 
have excluded
the very inner region of the cluster (r$<$20'', where the crowding 
conditions are most
severe) from our analysis.
In Fig. 3 the completeness factor $\phi=\frac{N_{rec}}{N_{sim}}$ 
(defined as the fraction of 
recovered stars over the 
total simulated ones) is plotted as a 
function of the $V_{555}$ 
magnitude in two different radial regions, namely
between 20'' and 60'' and at 
r$>$60'' from the cluster center, respectively.

\subsection{The RGB-Bump}

 The extended and populated RGB in 
 NGC 1978 gives the possibility to search for the so-called 
 RGB-Bump. This is the major evolutionary feature along the RGB. It flags 
 the point when the H-burning shell reaches the discontinuity in the 
 H-abundance profile left by the inner penetration of the convection.
 This feature has been predicted since the early theoretical models 
 \citep{iben} but observed for the first time in a globular cluster 
 almost two decades 
 later \citep{king}.  Since that first detection the RGB-Bump was  identified 
 in several GGCs \citep{fusi90, fer99, zoc99} and in a few galaxies
  in the Local Group
 (Sextant \citep{bel01}, Ursa Minor \citep{bel02}, Sagittarius \citep{monaco02}). 
 Accordingly with the prescriptions of \citet{fusi90}, we have used 
 the differential and 
 integrated luminosity function (LF) to identify the magnitude level 
 of the RGB-Bump in NGC 1978.  In doing this, we have {\it (1)} selected 
 stars belonging to the brightest ($V_{555}<$20.6) portion of the RGB;
 {\it (2)} carefully excluded the 
 bulk of the He-Clump and AGB stars
by eye; {\it (3)} defined the 
 fiducial ridge line for the RGB, rejecting those stars 
 lying at more than 2$\sigma$ from the ridge line.
Fig. 4 shows the final RGB sample (more than 600 stars)
and both 
 the differential  and 
integrated LFs.
The RGB-Bump appears in the differential LF as a well defined peak 
at $V_{555}^{bump}=19.10$ 
and it is confirmed in the integrated LF as a evident change in the slope. 

For both LFs the assumed bin-size is 0.1 mag; in order to check 
the uncertainty in the Bump magnitude level, we have tested the 
position of this feature by using LFs computed with different 
binning. The impact of the selected bin-size is not crucial: a 
difference of 0.2 mag corresponds
to a variation  $<$0.05 mag in the detection of RGB Bump.
By considering the intrinsecal width of the 
peak in differential LF, we estimate a conservative error 
$<$0.10 mag. 

 Finally, we note that the RGB Bump is brighter and reddest than
 of the bulk of the He-Clump and the latter merges into 
 the RGB at faintest magnitude
 ($V_{555}\sim$19.3, see Fig. 4; hence the possibility of contamination 
  is negligible.

\section{The cluster ellipticity}

Most globular clusters in the Galaxy show a nearly spherical shape, with a 
mean ellipticity 
\footnote{Note that ellipticity is defined here as $\epsilon$=1-(b/a),
where a and b represent major and minor axis of the ellipse, respectively.}
$\epsilon$=0.07 \citep{white} and more than 60\% with $\epsilon<$0.10.
One of the most remarkable exception is represented by $\omega$ Centauri
 that is clearly more
elliptical than the other GGCs: its ellipticity is 
$\epsilon$=0.15 in the external regions
with a evident decrease in the inner 
regions, with $\epsilon$=0.08 \citep{pancino}. 
Conversely, the LMC clusters (as well as those in the SMC)
show a stronger departure from the spherical simmetry. 
\citet{geisler} 
estimated the ellipticities of 25 popolous LMC clusters, finding a mean 
value of $\epsilon$=0.22;
\citet{goodwin} obtained a lower average value for the LMC clusters 
($\epsilon$=0.14), 
but still higher than the mean ellipticity of the GGCs. 
Moreover, 
in the LMC the presence of many double or triple 
globular clusters has been interpretated as a clue of the possibility of merger 
episodes between subclusters with the result to create stellar clusters  
with high ellipticities \citep{bathia}.\\
Previous determinations \citep{geisler, fischer} 
suggested large values of ellipticity for NGC 1978.

We have used the ACS catalog to derive a new measurement of the ellipticity of the cluster, 
in doing this we computed isodensity curves and adopting an adaptive kernel technique,
accordingly to the prescription of \citet{fuk}. In doing this we have adopted 
the center of gravity 
of the cluster computed using the near-infrared photometry obtained with SOFI \citep{m06}.
 The isodensity curves have been
computed using all the stars in the first chip with $V_{555}<$22 (approximately 
two magnitudes below the 
TO region) in order to minimize the incompleteness effects 
\footnote{Note that 
a different assumption on the magnitude threshold does not affect the result.}.
Finally, we have fitted the isodensity curves with ellipses. Fig. 5 shows
the cluster map with the isodensity contours (upper panel),
 the corresponding best fit ellipses 
(central panel) and their ellipticity as a function of the semi-major
 axis in arcsecond (lower panel). No evidence of subclustering or double
  nucleus is found. The average value of the ellipticity results $\epsilon$=0.30
(with a root mean square of 0.02), without any radial trend. 
This value is in good agreement with the previous estimates
\citep{geisler,fischer} and confirms the surprisingly high ellipticity of NGC 1978.

\section{The cluster age}

The determination of the age of a stellar population requires 
an accurate measure of the MS TO and the knowledge of the distance modulus, 
reddening and overall metallicity.
For NGC 1978 we used the recent accurate 
determination of [Fe/H]=-0.38$\pm0.02$ dex \citep{f06} and 
$<[\alpha/Fe]>$ almost solar (Mucciarelli et al., in preparation),
based on high-resolution spectra, to derive the overall metallicity [M/H]. 
In doing this, 
we adopted the relation presented by \citet{salaris}:
$$[M/H]\sim[Fe/H]+\log{(0.694\cdot10^{<[\alpha/Fe]>}+0.306)},$$
obtaining [M/H]$\sim$ -0.37 dex.\\
In the case of intermediate age stellar systems, the measurements of the age
is complicated by the presence of a convective core, whose size needs to be
parametrized ($\Lambda_{os}$)
\footnote{The overshooting efficiency is parametrized 
using the mixing length theory
\citep{bome} with $\Lambda_{os}$=1/$H_p$ 
(where $H_{p}$ is pressure scale height) that quantifies the overshoot distance {\sl above} 
the Schwarzschild border in units of the pressure scale height.
Some models as 
the "Padua ones" define this parameter as the overshoot distance {\sl across} 
the Schwarzschild border, hence the $\Lambda_{os}$ values from different models are not 
always directly comparable.
}. 
We then use different sets of theoretical isochrones 
with different input physics, in order to study the impact of the 
convective overshooting in reproducing the morphology of the main 
evolutionary sequences in the CMD.

\begin{itemize}
\item {\bf BaSTI models}: BaSTI ( {\sl A Bag of Stellar Tracks and 
Isochrones}) evolutionary code described in \citet{pietr}  computes 
isochrones with and without the inclusion 
of overshooting. The 
overshoot efficiency 
 depends on the stellar mass: (1) 
$\Lambda_{os}$=0.2 for masses larger than 1.7 $M_{\odot}$;
 (2) $\Lambda_{os}=0.25\cdot(M/M_{\odot}-0.9)$ 
 for stars in the 
1.1-1.7 $M_{\odot}$ range;
 (3) $\Lambda_{os}$=0  for stars less massive than 
1.1 $M_{\odot}$. \\

\item {\bf Pisa models} : PEL ({\sl Pisa Evolutionary Library}, 
\citet{cast03}) provides an
homogeneous set of isochrones computed without overshooting and with two
different values of $\Lambda_{os}$, namely 0.1 and 0.25. \\

\item {\bf Padua models} : in these isochrones \citep{gir00} 
 $\Lambda_{os}^{Padua}$=0 
for stars less massive than 
1 $M_{\odot}$, where the core is fully radiative.
The overshooting efficiency has been assumed to increase with 
stellar mass, according to the 
relation $\Lambda_{os}^{Padua}=M/M_{\odot}-1$ in the 1-1.5 $M_{\odot}$ 
range; 
above 1.5 $M_{\odot}$ 
a costant value of $\Lambda_{os}^{Padua}$=0.5 is assumed. 
Note that this value corresponds to $\Lambda_{os}\sim$ 0.25 
in the other models, 
where the extension 
of the convective region (beyond the classical boundary of 
the Schwarzschild 
criterion) is measured with respect to the convective core border.\\

\end{itemize}

\subsection{The morphology of the evolutionary sequences}

From each set of theoretical models, we selected  isochrones with Z=0.008 
(corresponding to [M/H]=-0.40 dex), consistent with the overall 
metallicity of the cluster
and we assumed a distance modulus $(m-M)_{0}\sim$18.5 
\citep{vdb98,clementini,alves} and 
E(B-V)=0.10 \citep{persson83}. However, in order to obtain the best fit to the
observed sequences with each isochrone set,
we left distance modulus and reddening to vary by $<\vert$10$\vert$\%  
and $<\vert$30$\vert$\% , respectively. 
Fig. 6 shows the best fit results for each isochrone set, 
while Table 1 lists the corresponding best fit values of age, 
reddening, distance modulus and 
the predicted magnitude level for the RGB-Bump.
 The best fit solution from each model set has been identified as the
  one matching the following
 features: ({\sl i}) the He-Clump magnitude level, ({\sl ii}) the magnitude
  difference between 
the He-Clump and the SGB and ({\sl iii}) the color extension of the SGB. 
The theoretical isochrones
have been reported into the observational plane by means of suitable transformations 
computed by using the code described in \citet{origlia} 
and convolving the model atmospheres by \citet{bcp} with the ACS filter responses.
In the following, we 
briefly discuss  the comparison between the observed evolutionary features and 
 theoretical predictions.

\begin{itemize}
\item {\bf BaSTI models}:
By selecting canonical models from the BaSTI dataset,
the best fit solution gives an age of 1.9 Gyr, with E(B-V)=0.09
 and a distance modulus 
of 18.47. Despite of the good matching of the He-Clump and SGB magnitude level, 
and the RGB slope, this isochrone does not properly reproduce the  
shape of the TO region and the 
{\it overall contraction phase } ({\it panel (a)} of Fig.6). The best-fit 
solution from overshooting models gives an age 
of 3.2 Gyr, $(m-M)_0$= 
18.43 and E(B-V)=0.09, and matches the main loci of the evolutionary sequences 
in the CMD. In particular, 
this isochrone provides a better match to the hook-like region (between the MS 
and the SGB, see {\it panel (b)} of Fig.6). \\

\item {\bf PEL models}: {\it Panels (c), (d)} and {\it (e)} of Fig. 6 
show the best fit solutions obtained by
selecting 3 different $\Lambda_{os}$. In all cases $(m-M)_0$=18.5 and 
E(B-V)=0.09 are used. As can be seen values of 
$\Lambda_{os}$=0 and $\Lambda_{os}$=0.25
isochrones fail to fit the SGB extension and the hook-like feature, conversely 
a very good fit is obtained with a mild-overshooting ($\Lambda_{os}$=0.1)
 and an age of $\tau$=1.9 Gyr.\\

\item {\bf Padua models}: 
The best-fit solution gives 
$\tau$=2.2 Gyr, $(m-M)_0$=18.38 and 
E(B-V)=0.07 ( {\it Panels (f)} of Fig. 6). 
This isochrone well-reproduces the complex structure of the TO 
and the core contraction stage, as well as the SGB structure 
and the RGB slope. 
However, it requires distance modulus and reddening significantly lower 
than those generally adopted for the LMC.
\end{itemize}
From this comparison, it turns out that  
only  models with overshooting are able to best fit  
the morphology of the main evolutionary sequences in the observed CMD. 
In particular, the best fit solutions have been obtained with 
the BaSTI overshooting model with $\tau$=3.2 Gyr, 
the PEL mild-overshooting model ($\Lambda_{os}$=0.1) and $\tau$=1.9 
Gyr and the Padua model with $\Lambda_{os}^{Padua}$=0.25 
(corresponding to $\Lambda_{os}$=0.25) and $\tau$=2.2 Gyr.

However, it must be noted that none of these models satisfactorly 
can fit the observed Bump level, the BaSTI and Padua models
 being $\approx$0.1 and 
0.3 mag fainter, respectively, and the PEL model $\approx$0.2 mag brighter, 
perhaps suggesting that evolutionary tracks for stars with M$>$1$M_{\odot}$ 
still need some fine tuning to properly reproduce the luminosity of this 
feature.

\subsection{Population ratios}
Since the comparison between the observed CMD and theoretical 
isochrones is somewhat qualitative,
we also performed a quantitative comparison
between theoretical and empirical population ratios:
this yields a direct check 
of the evolutionary timescales.
To do this, we define four boxes selecting the stellar 
population along the main evolutionary features in 
our CMD, namely the He-Clump, the SGB, the RGB
(from the base  
up to $V_{555}\sim$19.4) and finally 
the brightest ($\sim$ 1 mag) portion of the MS; these boxes 
are show in Fig. 7, overplotted to the cluster CMD. Star counts in
each box have been corrected for incompleteness, by dividing the observed 
counts by the $\phi$ factor obtained from 
the procedure described in Sec. 2.3 (see also Fig. 3) for each 
bin of magnitude.
 
Star counts have been also corrected for field contamination.
To estimate the degree of contamination by foreground and background 
stars we have applied a statistical technique.
We have used the CMD shown in Fig. 2 as representative of 
the field population.
The number of stars counted in each box in the control field
have been normalized to the cluster sampled area and finally subtracted 
from the cluster star counts. 
The {\sl final} star counts per magnitude bin in each
box have been estimated accordingly to the following formula:
$$N_{corr}=\frac{N_{obs}}{\phi}-N_{field},$$
where ${N_{obs}}$ are the observed counts and $N_{field}$ 
the expected field star counts.
We find $N_{MS}$=4331, $N_{SGB}$=632, $N_{RGB}$=450 and $N_{He_{Cl}}$=311, 
where $N_{MS}$, $N_{RGB}$, $N_{SGB}$ and $N_{He-Cl}$ are the 
number of stars in the box as 
sampling the MS, the RGB, the SGB and the He-Clump population, respectively.

Uncertainties in the computed population ratios have been estimated
 using the following formula 
$$\sigma_{R}=\frac{\sqrt{R^{2}\cdot\sigma_{D}^{2}+\sigma_{N}^{2}}}{D}$$
where $R=N/D$ is a given population ratio, N is the numerator and 
D the denominator of the ratio. 
The errors $\sigma_{N}$ and $\sigma_{D}$ for any population
have been assumed to
follow a Poisson statistics. In addition, in the error budget we also include 
the uncertainty due to the  
positioning of the box edges: note that a slightly different($\pm1\sigma$) 
assumption in the 
definition of the box edge has  little impact (typically 
7-8\%) on the star counts. This uncertainty has 
been quadratically added to the Poissonian error.

On the basis of the boxes shown in Fig. 7 we defined four population ratios 
\footnote{Note that the bluest portion of the SGB can be affected by blending. 
To check this effect, 
we also defined a second box sampling the SGB population ($SGB_s$), by excluding the
 bluest region at $(V_{555}-I_{814})<$0.7. The population ratios obtained by using this 
 selection box (and reported in Table 2) are fully consistent with the results
  by using the standard SGB box, suggesting that blending 
 effects (if any) in the SGB population have a negligible impact on the results.}, 
as listed in Table 2:
({\bf i}) $N_{RGB}$/$N_{SGB}$;
({\bf ii}) $N_{RGB}$/$N_{He-Cl}$;
({\bf iii}) $N_{SGB}$/$N_{He-Cl}$;
({\bf iv}) $N_{MS}$/$N_{(SGB+RGB)}.$ 
For each selected model, corresponding theoretical population 
ratios have been estimated by 
convolving the isochrone set shown in Fig.4 with an Initial
 Mass Function (IMF),
 according with 
the prescriptions of \citet{straniero91}.  In order to check 
the sensitivity of the
population ratios to the adopted IMF, we have used three 
different values for the IMF slope {\sl $\alpha$}: 2.35 \citep{salpeter}, 
2.30 \citep{kroupa} and 
3.5 \citep{scalo} at M$>$1$M_{\odot}$. In the considered mass range 
(between 1 and 2 $M_{\odot}$),
 the theoretical population ratios are poorly dependent on 
the assumed IMF, with a 16\% maximum 
variation (between Scalo and Kroupa IMFs) for the 
$N_{MS}$/$N_{(SGB+RGB)}$ ratios.
 Hence in the following the population ratios are computed
  by using a Salpeter IMF.

As results, we found that BaSTI and PEL canonical models predict a lower 
(by $<$40\%) 
of the $N_{MS}$/$N_{(SGB+RGB)}$ and higher 
 $N_{RGB}$/$N_{He-Cl}$ and $N_{SGB}$/$N_{He-Cl}$ 
 (by $<$35\% and $<$100\%, respectively) 
 population ratios with respect 
 to the observed ones. 
 Isochrones 
with high overshooting ($\Lambda_{os}$=0.2-0.25) show an opposite trend, 
with higher (by $<$50\%) $N_{MS}$/$N_{(SGB+RGB)}$  and 
lower (by $<$30\%) $N_{RGB}$/$N_{He-Cl}$ and $N_{SGB}$/$N_{He-Cl}$ ratios.
 The isochrone with $\Lambda_{os}$=0.1 from PEL dataset
reasonably reproduces all the
 population  ratios. Only  the $N_{MS}$/$N_{(SGB+RGB)}$ 
 ratio turns out to be $\sim$15\% lower 
 than the observed one.\\
We conclude that the best agreement with observations (both in
 terms of evolutionary sequence 
morphology and star counts) has been obtained by using PEL models
 computed
with a mild overshooting ($\Lambda_{os}$=0.1) and $\tau$=1.9 Gyr.
Also, the required values of distance modulus and reddening are
 fully consistent with 
those generally adopted for NGC 1978.
In order to estimate the overall age uncertainty, we took into 
account the major error source, 
namely the distance modulus. 
Hence, 
we have repeated the best-fitting procedure by using 
the PEL isochrones with mild-overshooting,
and varying the distance modulus by $\pm$0.05 and $\pm$0.1 mag 
with respect to the reference 
value of 18.5. 
A variation of $\pm$0.05 mag still allows a good fit of the CMD features with 
isochrones within $\mp$0.1 Gyr from the reference value of 1.9 Gyr. 
A variation of $\pm$0.1 mag in the distance modulus, does not allows 
to simultaneously fit the He-Clump magnitude level and the 
extension of SGB, whatever 
age is selected.
Hence, we can assign a formal error of $\pm$0.1 Gyr to our age estimate.

\section{Discussion and Conclusions}

The photometric analysis of the ACS-WFC CMD of NGC 1978
presented here provided three major results, that can be 
summarized as follows: 
{\it i)}~the firm detection of the RGB Bump 
at $V_{555}$=19.10$\pm$0.10,  
{\it ii)}~a new, independent 
estimate of the cluster ellipticity ($\epsilon$=0.30$\pm$0.02)
and {\it iii)}~an accurate measure of the cluster
 age ($\tau$=1.9$\pm$0.1 Gyr).

The detection presented here is the first clearcut detection 
of the RGB Bump in an intermediate age cluster and it confirms 
the theoretical expectation that this feature also occurs 
in relatively massive stars (the estimated TO mass for this cluster 
is $\sim$1.5 $M_{\odot}$, see Tab. 1). Note that this result opens 
the possibility to study the behaviour of the RGB Bump as a function 
of the cluster age and it provides crucial insight on the 
internal structure of intermediate mass stars.

The high-ellipticity ($\epsilon$=0.30$\pm$0.02) of NGC 1978 poses two 
major questions: {\it i)}~why the LMC clusters in general, and NGC 1978 
in particular, are, in average, more elliptical than those in the Milky Way?
{\sl ii)}~why NGC 1978 is more elliptical than the other LMC clusters?\\
\citet{goodwin} suggests that the relatively small LMC tidal field 
can preserve the pristine triaxial structure of the clusters, 
while the strong tidal field of our Galaxy tend to destroy it,
thus removing at least part of the ellipticity.\\ 
In order to explain the especially high ellipticity of NGC 1978 
three main hypothesis 
have been proposed in the past: a merging episode, a rotation effect
and an anisotropic velocity dispersion tensor \citep{fischer}. 
The merging scenario has been proposed 
because the broad RGB from ground-based BVRI 
photometry \citep{alcaino} and because the preliminary 
evidence of a metallicity dispersion from 
high-resolution spectroscopy of 
two RGB stars \citep{hill00} ($\delta$[Fe/H]$\sim$0.2-0.3 dex). 
However, the tiny RGB sequence presented in this work as well as 
the recent iron abundance estimate from  high-resolution spectra 
of eleven RGB stars presented by \citet{f06}
definitely excluded any significant metallicity spread within 
the cluster.

Our new age estimate ($\tau$=1.9$\pm$0.1 Gyr) 
of NGC 1978, coupled with the new 
iron abundance determination ([Fe/H]=-0.38$\pm$0.02 dex) 
by \citet{f06}, provide new 
coordinates for this cluster in the age-metallicity plane. 
This is especially important, since 
the correct shape of the AMR in the LMC is still
matter of debate:
in particular the origin of the observed bimodality in the LMC 
cluster age distribution
has been interpreted as the evidence for two major episodes of star formation.
\citet{pagel} computed two different AMR semi-empirical models for the LMC,
with a continuous star formation and 
with two burst episodes occurred $\sim$14 and 3 Gyr ago, respectively.
Fig. 8 shows the results of these theoretical predictions.
For comparison, the 
position of NGC 1978 in the age-metallicity diagram based on 
{\it i})~old metallicity ($\rm-0.6<[Fe/H]<-0.4$
\citet[see ][]{ols91,defre}) and 
age \citep[2-3.3 Gyr][]{ols84,geisler97,gir95} estimates (grey box),
{\it ii})~the iron abundance  
by \citet{hill00} and age by \citet{bomans} (open square), 
and finally 
{\it iii})~the most recent 
metallicity by \citet{f06}
and age from the present work (big black dot) are shown. 
It is interesting to note that old coordinates (grey box) barely 
fit with the bursting scenario, 
while the more recent measurements by \citet{hill00} place the cluster 
far below any model.  
Our new coordinates are somewhat consistent with both the 
proposed star formation scenarios.\\ 
Similar accurate 
([Fe/H], $\tau$) coordinates for a
 significant number of LMC clusters with different ages and 
 metallicities are urgently needed to disentagle 
different formation scenarios. This is the aim of our ongoing global 
project. By combining detailed chemical abundance
(from high-resolution spectra) and ages (from high quality photometry) 
to a number of pillar LMC clusters, we plan to calibrate a suitable 
age and metallicity scale for the entire LMC globular cluster system, 
with the ultimate goal of providing a robust AMR.

\acknowledgements  
We warmly thank the anonymous referee for his/her suggestions.
This research 
was supported by the Agenzia Spaziale Italiana (ASI) and the 
Ministero dell'Istruzione, del\-l'Uni\-versit\`a e della Ricerca.

\begin{deluxetable}{lcccccc}
\tablecolumns{7} 
\tablewidth{0pc}  
\tablecaption{Age, distance modulus, reddening , Turn-Off mass 
and magnitude level of the RGB-Bump
from best-fit BaSTI, PEL and Padua isochrones.}
\tablehead{ 
\colhead{}& \colhead{BaSTI}  & \colhead{BaSTI}& \colhead{PEL}
 &  \colhead{PEL} & 
\colhead{PEL} & \colhead{PADUA}  \\
& \colhead{$\Lambda_{os}$=0} &\colhead{$\Lambda_{os}$=0.2}& \colhead{$\Lambda_{os}$=0}
 &  \colhead{$\Lambda_{os}$=0.1} & 
\colhead{$\Lambda_{os}$=0.25} &   \colhead{$\Lambda_{os}$=0.25}    }
\startdata 
  Age (Gyr)               &  1.9 & 3.2   & 1.7 & 1.9   & 2.5  & 2.2  	    \\
  $(m-M)_0$               &  18.47 & 18.43   & 18.50 &  18.50  & 18.50  & 18.38    \\
  E(B-V)                  &  0.09 &  0.09  & 0.09 &  0.09  & 0.09  & 0.07  \\
  $M_{TO}$ ($M_{\odot}$)  &  1.47 & 1.45   & 1.49 & 1.49   & 1.44  & 1.45 	    \\
  $V_{555}^{Bump}$        &  19.10 & 19.22 & 18.73 & 18.88 & 19.39 & 19.44 \\
  \enddata 
\end{deluxetable}

\begin{deluxetable}{lccccccc}
\tablecolumns{8} 
\tablewidth{0pc}  
\tablecaption{Theoretical population ratios from BaSTI, PEL and Padua best-fit isochrones, and 
corresponding observed ratios for NGC 1978.}
\tablehead{ 
\colhead{~~~Population Ratio~~~}& \colhead{BaSTI}  & \colhead{BaSTI}& \colhead{PEL}
 &  \colhead{PEL} & 
\colhead{PEL} & \colhead{PADUA} &\colhead{Observed} \\
& \colhead{$\Lambda_{os}$=0} &\colhead{$\Lambda_{os}$=0.2}& \colhead{$\Lambda_{os}$=0}
 &  \colhead{$\Lambda_{os}$=0.1} & 
\colhead{$\Lambda_{os}$=0.25} &   \colhead{$\Lambda_{os}$=0.25}    }
\startdata 
   MS/(RGB+SGB) & 2.40  &	6.05	  & 2.10 & 3.27   & 7.06  &  6.92 & 4.00$\pm$0.40	\\
   SGB/He-Cl    & 4.16  &	1.63	  & 4.90 & 2.01   & 1.62  &  1.05 & 2.03$\pm$0.22 \\
   RGB/He-Cl    & 1.95  &	1.07	  & 2.41 & 1.64   & 1.03  &  0.78 & 1.45$\pm$0.19 \\
   RGB/SGB      & 0.58  &	0.65	  & 0.49 & 0.81   & 0.64  &  0.75 & 0.71$\pm$0.10	\\
   $SGB_s$/He-Cl & 1.39 &       0.64      & 1.43 & 0.85   & 0.61  &  0.36 & 1.12$\pm$0.12 \\
   MS/(RGB+$SGB_s$)& 4.14 &     9.61      & 4.00 & 4.75   & 11.43 &  11.29 & 5.43$\pm$0.42 \\
  \enddata 
\end{deluxetable}

\begin{figure}[h]
\plotone{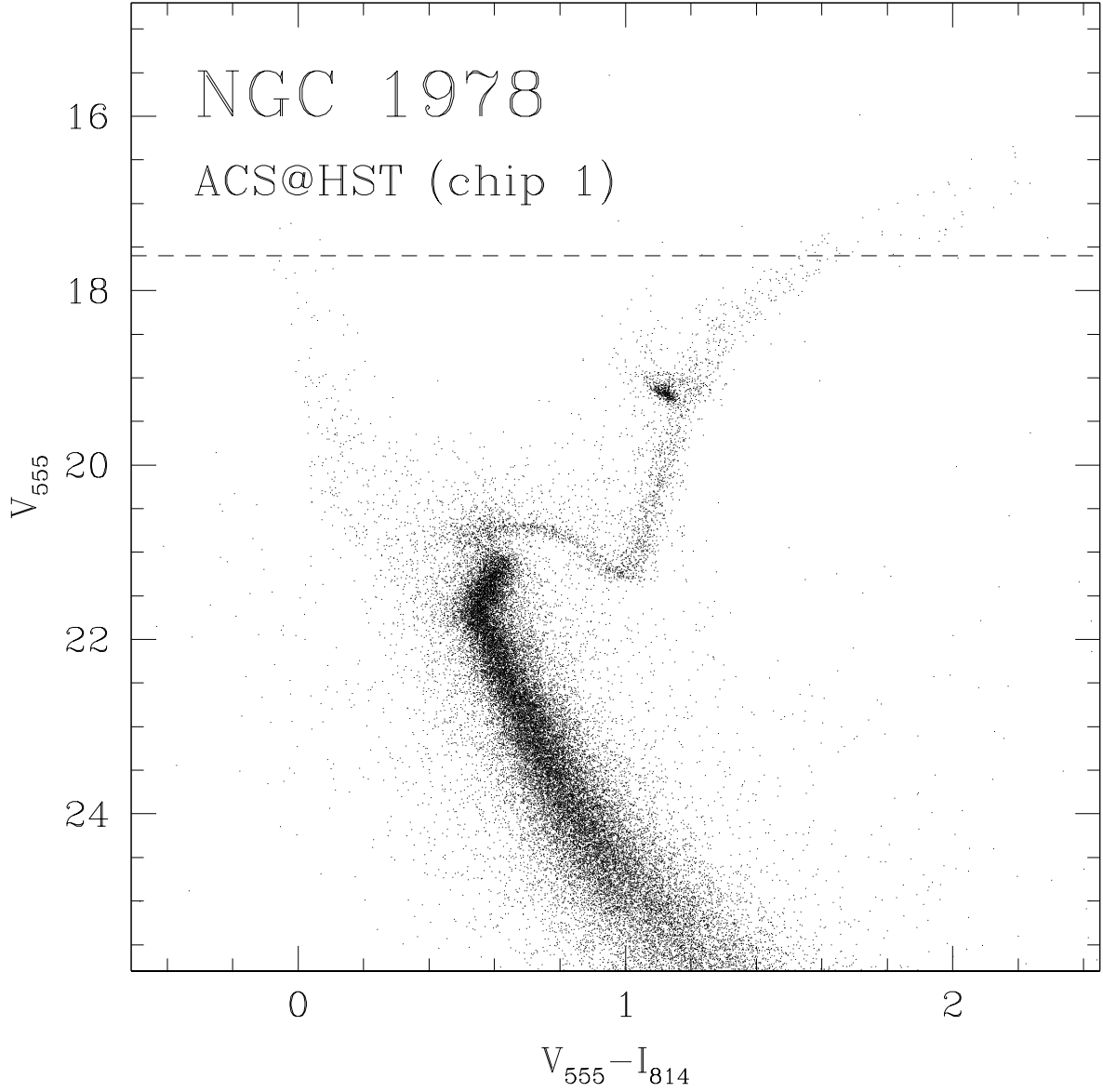}
\caption{(V, V-I) CMD of the LMC cluster NGC 1978, obtained with ACS@HST (only first chip).
The dashed line indicates the saturation level.}
\label{}
\end{figure}
 
\begin{figure}[h]
\plotone{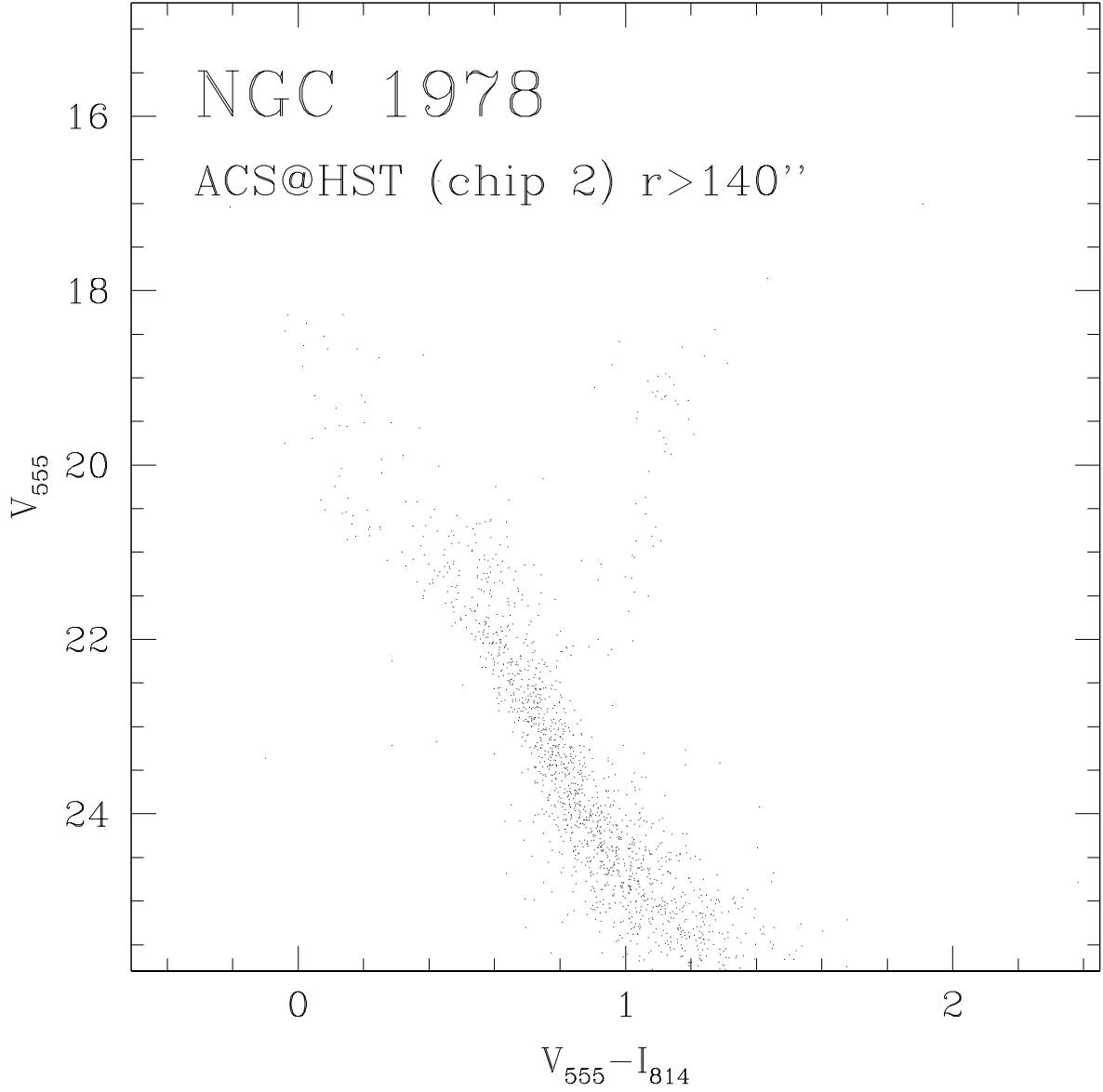}
\caption{(V, V-I) CMD of the outer region (r$>$140'' from the cluster center) of NGC 1978, 
as obtained with ACS@HST (only second chip).}
\label{}
\end{figure}
 
\begin{figure}[h]
\plotone{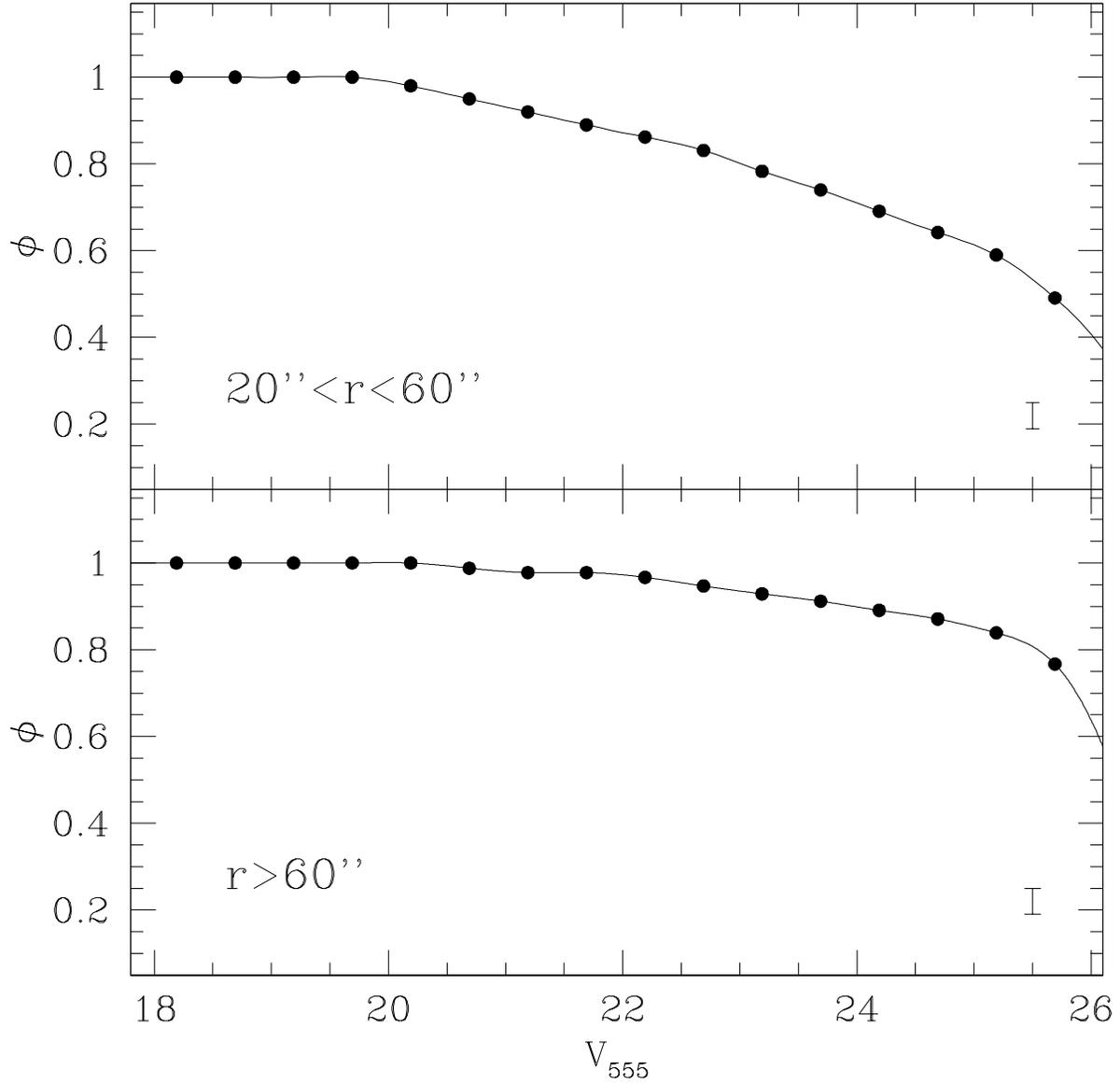}
\caption{Completeness curves computed in two radial sub-regions of NGC 1978.
 The black points indicate the value of the $\phi=\frac{N_{rec}}{N_{sim}}$ 
 parameter calculated for each 
magnitude bin. The completeness curves have been 
computed for $V_{555}<$17.6, corresponding to 
the saturation level. 
Tipycal errorbars are also indicated.}
\label{}
\end{figure}

\begin{figure}[h]
\plotone{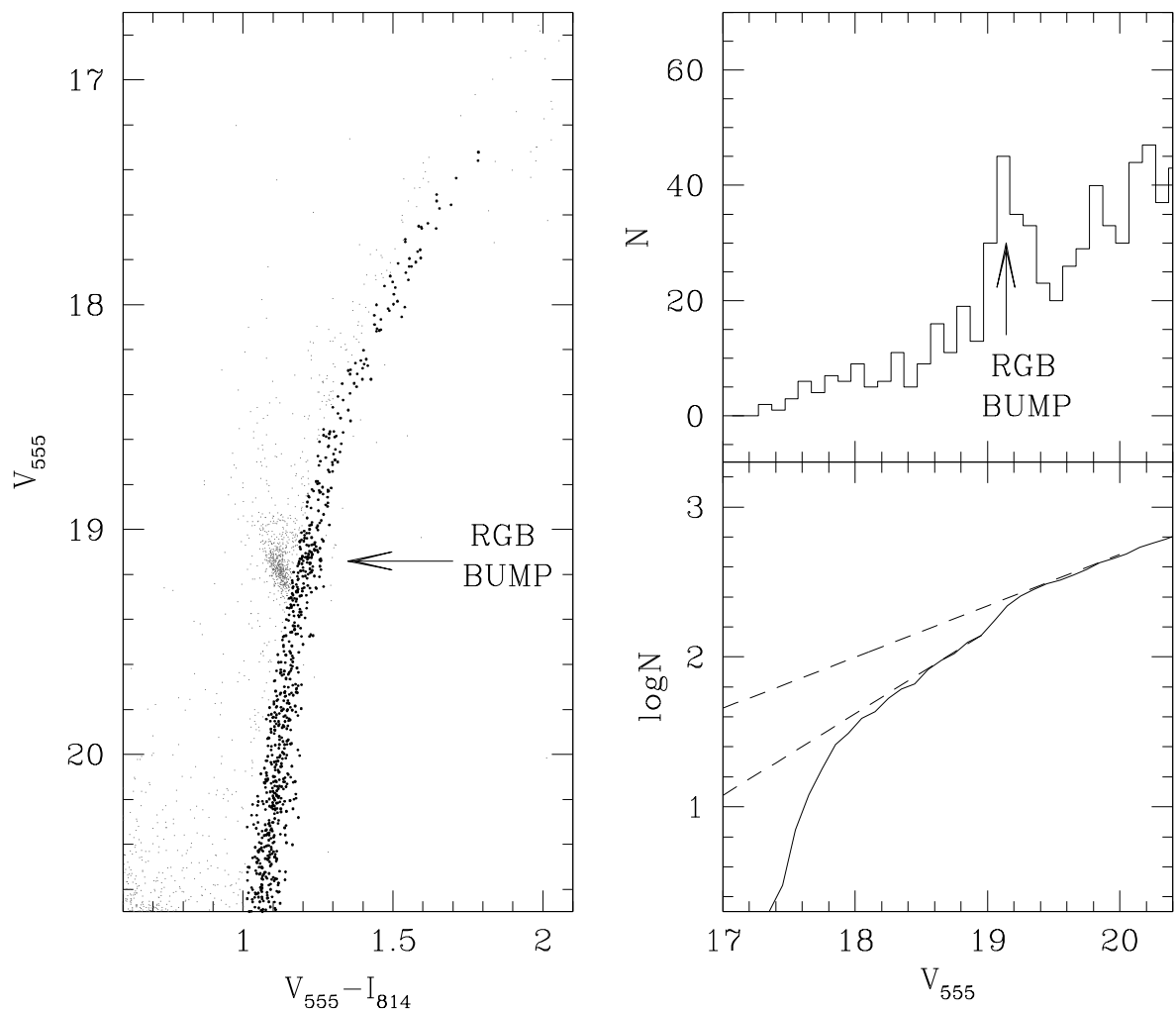}
\caption{Left: the bright portion of the CMD of NGC 1978 (grey points) with the 
selected RGB stars (black points). The arrow indicates the magnitude of the RGB Bump.
Right:
differential (upper panel) and integrated (lower panel) LFs, computed for 
the RGB stars, excluding the He-Clump and AGB populations. The arrow in the 
upper panel indicates the position of the RGB Bump. The dashed lines in the lower panel 
are the linear fit to the regions above and below the RGB Bump.}
\label{}
\end{figure}

\begin{figure}[h]
\plotone{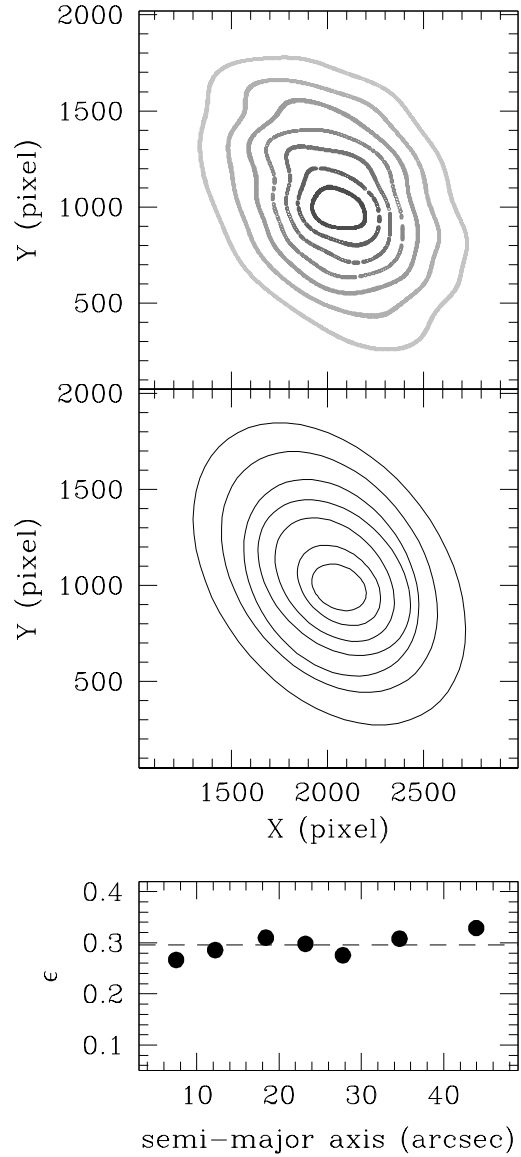}
\caption{Upper panel: the map of NGC 1978 with the isodensity contours; central panel: 
the best fit ellipses of the isodensity contours; lower panel: ellipticity of 
the best fit ellipses as a function of the semi-major axis in arcsecond. The 
horizontal dashed line indicate the mean value.}
\label{}
\end{figure}

\begin{figure}[h]
\plotone{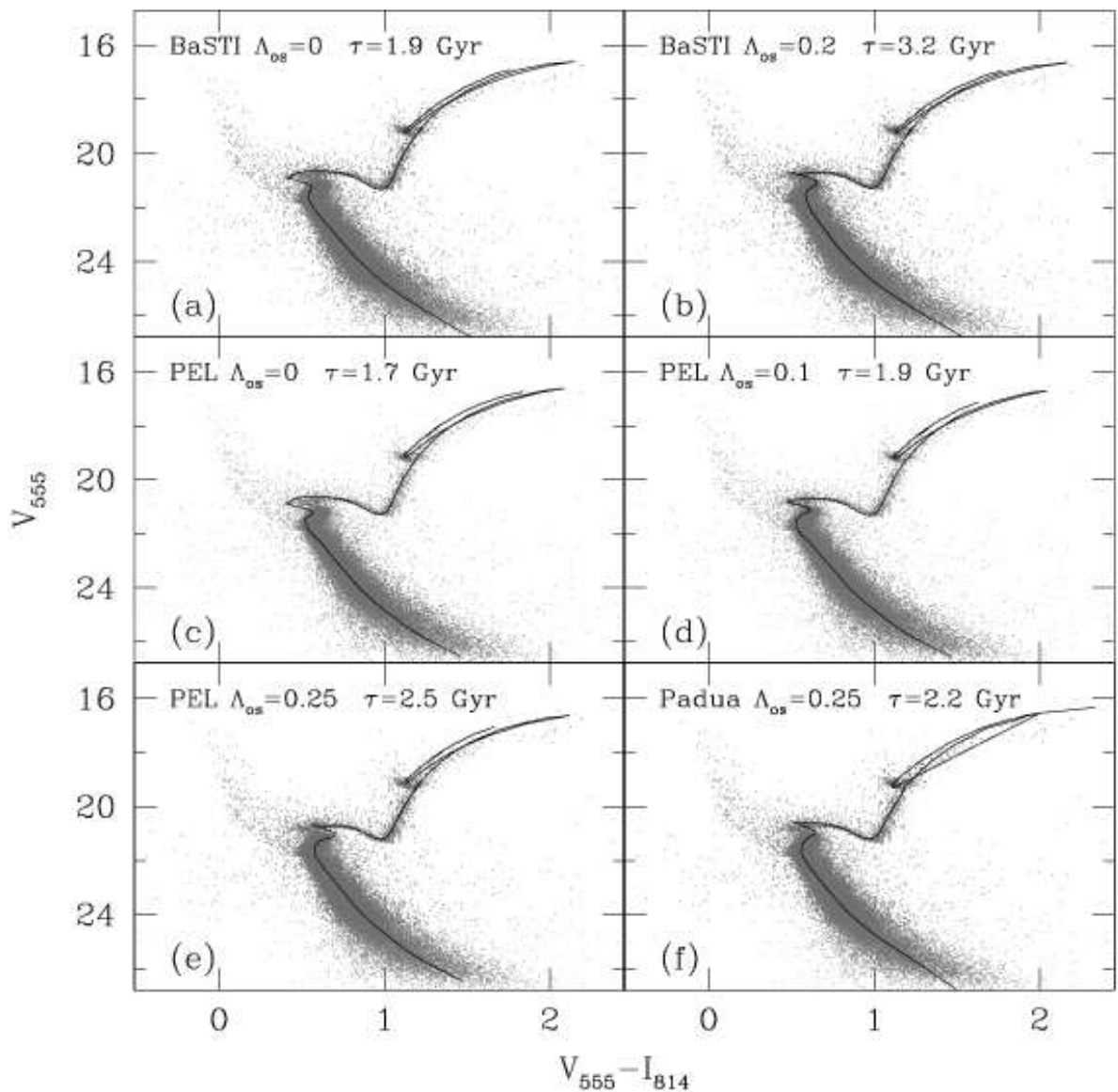}
\caption{Best-fit theoretical isochrones overplotted to the observed CMD of NGC 1978 (only
stars at r$>$20'' from the cluster center are  plotted) obtained with 
theoretical isochrones: each panel shows a different model and the corresponding 
$\Lambda_{os}$ value and age. Reddening and 
distance moduli for each model are reported in Table 1.}
\label{}
\end{figure}

\begin{figure}[h]
\plotone{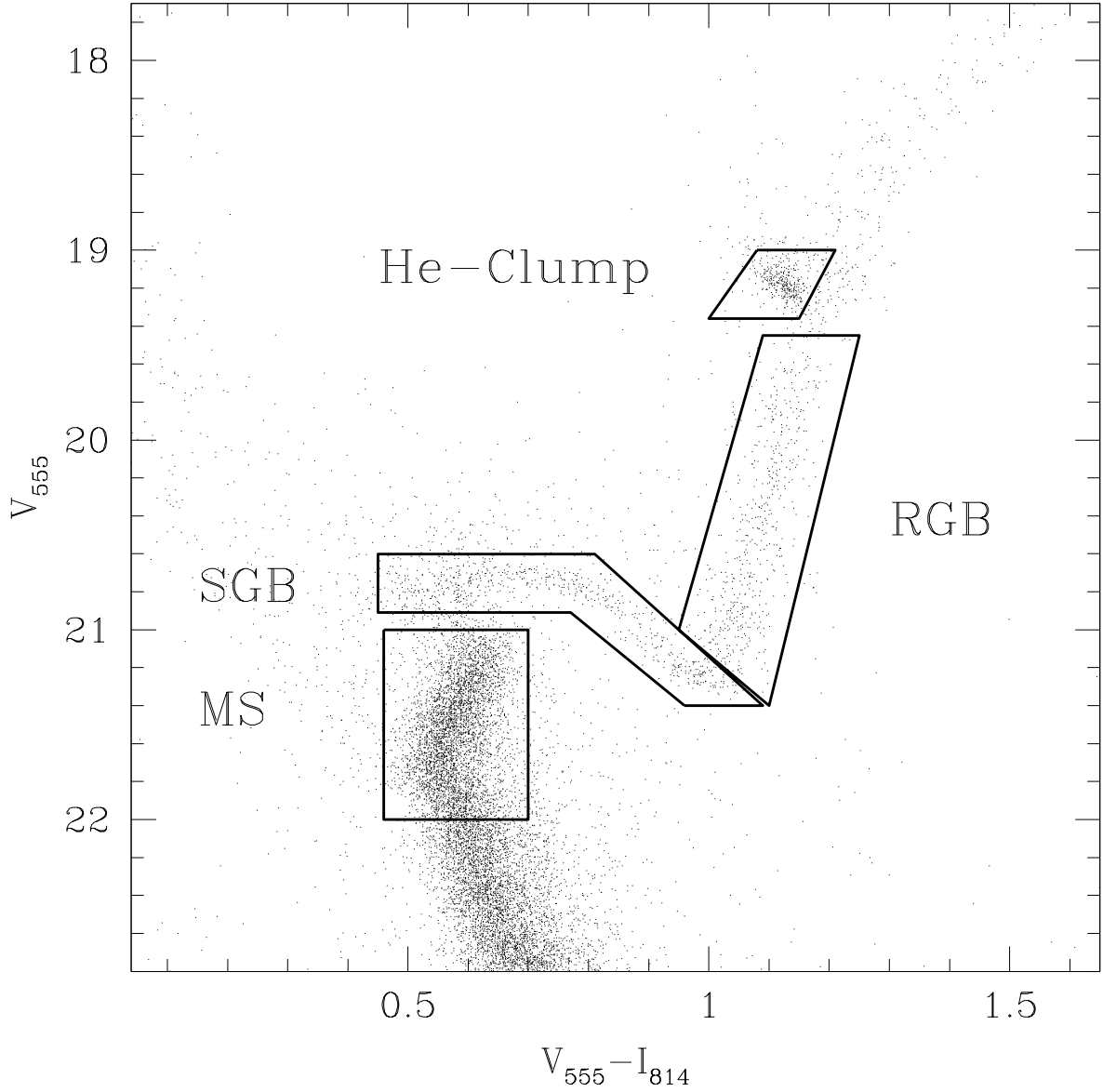}
\caption{The bright portion of the NGC 1978 CMD (only
stars at r$>$20'' from the cluster center are  plotted)
with the selection boxes adopted to sample the MS, SGB, RGB and He-Clump populations.}
\label{}
\end{figure}

\begin{figure}[h]
\plotone{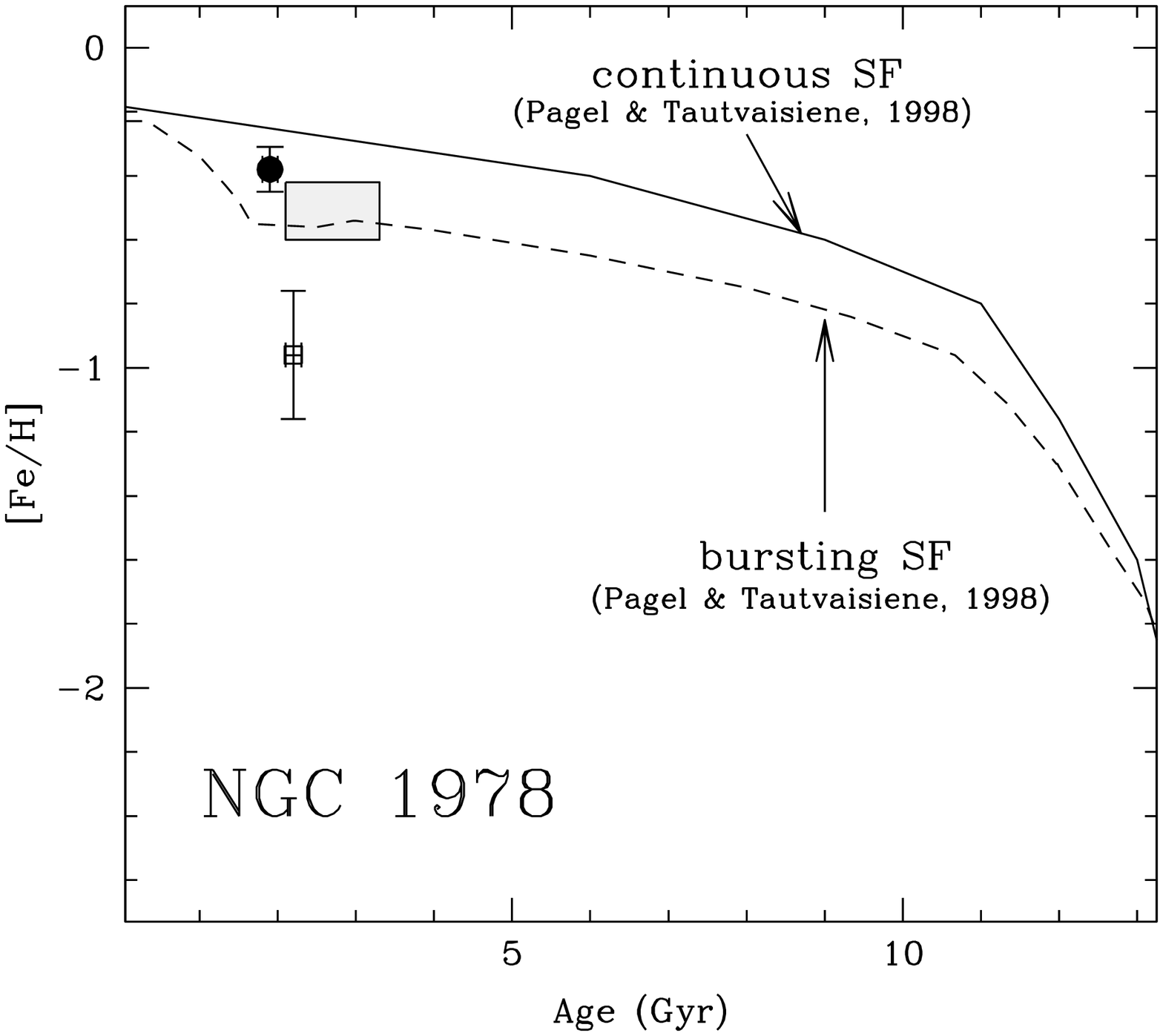}
\caption{Theoretical predictions for the LMC AMR computed by 
\citet{pagel}: continuos line refers to a AMR obtained assuming a continuous SF, 
and dashed line corresponding to a bursting model. The black point indicates the 
position of NGC 1978 using the metallicity from \citet{f06} and the age derived in this study. 
The open square refers the position for this cluster using metallicity by \citet{hill00} and age by \citet{bomans}. 
The grey box shows previous metallicity ($\rm-0.6<[Fe/H]<-0.4$) and 
age (2-3.3 Gyr) estimates from low resolution data.}
\label{}
\end{figure}

\end{document}